\pdfoutput=1
\documentclass[submission]{eptcs}
\providecommand{\event}{ACT 2019}

\usepackage[utf8]{inputenc}
\usepackage[T1]{fontenc}
\usepackage{graphicx}
\usepackage{tabularx, ragged2e, booktabs}
\usepackage{amsthm} %
\usepackage{mathtools} %
\usepackage{amssymb} %
\usepackage{bbold} %
\usepackage{subcaption} %
\input{idrisLang.sty}
\usepackage[
	style=numeric,
	maxnames=3, 
	bibencoding=utf8,
	backend=biber,
	defernumbers=true,
	backref=true,
	hyperref=true,
	url=true,
	isbn=false,
	alldates=long
]{biblatex}
\DeclareSourcemap{
  \maps[datatype=bibtex]{
    \map{
      \step[fieldsource=archiveprefix,final]
      \step[fieldset=url,null]
    }  
    \map{
      \step[fieldsource=doi,final]
      \step[fieldset=url,null]
    }  
  }
}
\makeatletter
\let\blx@rerun@biber\relax
\makeatother

\AtEveryBibitem{%
	\clearlist{address}
	\clearlist{location}
	\clearfield{series}
}
\usepackage{hyperref} %

 \AfterEndEnvironment{theorem}{\noindent\ignorespaces}

 \AfterEndEnvironment{proposition}{\noindent\ignorespaces}

 \AfterEndEnvironment{definition}{\noindent\ignorespaces}

 \AfterEndEnvironment{fact}{\noindent\ignorespaces}

 \AfterEndEnvironment{example}{\noindent\ignorespaces}
\AfterEndEnvironment{proof}{\noindent\ignorespaces}
\newcommand{\idrisct}{\texttt{idris-ct}\xspace}

\title{\texttt{idris-ct}: A Library to do Category Theory in Idris}
\author{
	Fabrizio Genovese
	\email{\quad 0000-0001-7792-1375\quad }
	\and
	Alex Gryzlov
	\email{\quad 0000-0001-6188-0417\quad }
  \and
	Jelle Herold
	\email{\quad 0000-0002-1966-2536\quad }
	\and
	Andre Knispel
  \email{\phantom{\quad 0000-0000-0000-0000\quad }}
  \and
	Marco Perone
	\email{\quad 0000-0002-1004-0431\quad }
  \and
	Erik Post
  \email{\quad 0000-0002-8111-9593\quad }
  \and
	Andr\'e Videla
  \email{\quad 0000-0002-7298-6230\quad }\\\\
  Statebox Team
	\email{research@statebox.io}
  }

\def\titlerunning{\texttt{idris-ct}: A Library to do Category Theory in Idris}
\def\authorrunning{Genovese, Gryzlov, Herold, Knispel, Perone, Post, Videla}

\begin{document}
\maketitle
\begin{abstract}
\noindent
We introduce \idrisct, a Idris library providing verified type 
definitions of categorical concepts. \idrisct strives to be a bridge between
academy and industry, catering both to category theorists who want to implement
and try their ideas in a practical environment and to businesses and engineers
who care about formalization with category theory: 
It is inspired by similar libraries developed for theorem proving but 
remains very practical, being aimed at  software 
production in business. Nevertheless, the use of dependent types allows 
for a formally correct implementation of categorical concepts, so that 
guarantees can be made on software properties.
\end{abstract}
\section{Introduction and motivation}
The Statebox project~\cite{StateboxTeam2017} is centered around the 
idea of building a graphical programming language using the 
well-known correspondence between Petri nets and free symmetric 
monoidal categories~\cite{StateboxTeam2019a}. Trying to turn this 
idea into practice, our team was confronted with the problem that many 
research results in the field were strictly 
theoretic -- e.g.~\cite{Sassone1995, Baldan2003} -- with 
implementation never to be considered a priority. The effort to 
overcome these issues led to some novel 
research -- e.g.~\cite{Genovese2019a, Genovese2019} -- but more 
importantly to the development of some basic computational tools 
needed to carry out the job, as the one we are going to introduce here. 

This paper has the purpose of announcing and describing \idrisct, 
which is open to external contributions, to the ACT community: 
We are already receiving pull requests, and would very much 
appreciate an involvement of the ACT community in this open project. 
Here our scope will remain broad, and we will explain some design 
choices at the expense of in-depth technical specification: For instance, 
the examples we provide may look trivial to expert programmers and researchers.
We redirect them to the \idrisct Github repository~\cite{StateboxTeam2019}, where 
everything is worked out and documented in a greater level of detail, 
and more involved concepts are covered.
\section{What is \idrisct?}
\idrisct is an Idris library focused on providing verified type 
definitions of important concepts in category theory, such as 
``category'', ``functor'', ``monoidal category'', etc. The project is 
open source, licensed under AGPL 3.0~\cite{AGPL3} (license 
exemptions are available), and its source code can be found on 
Github~\cite{StateboxTeam2019}. Installation instructions are available 
on the repository.

The original requirements laid down for \idrisct focused on obtaining 
a library practical enough to be production-ready in software 
development for business and enterprise. In this respect, pre-existing 
solutions -- see e.g.~\cite{Peebles} -- did not fit our requirements.
Indeed, we wanted a codebase that allowed us to answer questions 
such as ``What kind of runtime costs are we looking at'' and ``What 
would be a more ergonomic API for the users of the library''. Similar 
goals have been set by Rydeheard and Burstall in~\cite{Rydeheard1988} but, as they 
themselves declare, ``we hoped that [this work] would provide a tool 
for advanced programming, harnessing the abstraction of category 
theory for use in program design. These hopes have not really been 
realized by our work so far.'' As a result, we felt compelled to develop 
an alternative solution.
\subsection{Why Idris}
One might ask why we chose to use Idris for our coding endeavours. 
The main reason is that one of our priorities 
is \emph{correctness}: We want to implement the categorical 
connection between nets and symmetric monoidal categories 
\emph{formally}.

In order to achieve this goal, we want to employ a programming 
language that is powerful enough to let us \emph{prove} properties 
about our code. For example, we want to implement structures so 
that defining a category in our model is equivalent to 
\emph{mathematically proving} that it is indeed a category in 
the ``traditional''~\cite{MacLane1978} sense. 

At the same time we need a language backed by a strong community, 
well versed in the problems faced with software \emph{engineering}. 
Its ecosystem should provide HTTP and parsing libraries, C-FFI, 
filesystem access and web capabilities. In short, we require a very 
specific blend of formal methods and pragmatic solutions.

There exist multiple popular functional programming 
languages allowing different degrees of formalism and pragmatism, 
such as Coq~\cite{TheCoqConsortium}, Haskell~\cite{Haskell.org}, 
Idris~\cite{Brady2013, Brady2017} and Agda~\cite{agda, Norell2007}. 
We decided to go for Idris because we strongly believe in its goal to be a 
``Pacman-complete''~\footnote{
  \emph{Pacman-completeness} 
  is a folklore concept in programming communities representing the 
  idea that having a powerful language is not enough for it to be useful. 
  For instance, PowerPoint presentations are 
  Turing-complete~\cite{Wildenhain2017}, and yet we are not building 
  operating systems with them. Pacman-completeness serves the idea 
  that a programming language is ``useful enough'' so that you can use 
  it to implement the game ``Pacman'' without having a seizure.
}
programming language first, and a theorem prover second. 
This design choice makes an important difference, as it is highlighted by
practical projects as~\cite{Fowler2014} and~\cite{McKenna}.
Additionally, Idris' successor Blodwen~\cite{Brady} already shows 
promise by featuring linear types and a more efficient compiler.

It is worth noting that our design choices are, at the 
moment, quite unique in the applied category theory landscape: Other 
projects, like~\cite{Reutter2019},~\cite{Rydeheard1988} and~\cite{ConexusTeam}, are 
implemented using languages that are either imperative or have a 
weaker type system, hence coming with less formal guarantees 
(e.g. composing morphisms using the algorithms in~~\cite{Rydeheard1988} can throw 
exceptions). Conversely, many projects striving to implement 
categorical gadgets in a formally verified way are confined to the realm 
of theorem proving (e.g.~\cite{Peebles, Gross2014}), with no 
plans -- to our knowledge -- to provide the needed integrations to make 
them usable in industry.
\subsection{Our Idris is literate}
Another noteworthy design decision is to write \idrisct in 
\emph{literate} Idris. Literate programming is a paradigm where, 
instead of prepending a symbol to instruct the compiler that a given 
line is a comment (such as \texttt{\%} in \LaTeX), the user does the 
opposite, prepending a symbol to the lines that \emph{do not} have to 
be ignored by the compiler.

This programming mode allows for having extensive blocks of text 
documenting code, and to present the code itself as a story 
(hence the ``literate''). In our case, \idrisct can be compiled both to 
Idris, obtaining a compiled source in the usual sense, or to \LaTeX, 
obtaining a pdf explaining the code in detail.

The choice of documenting our project with literate Idris stems from 
the fact that \idrisct is not aimed exclusively at academia. In fact, we believe that
many developers in industry use concepts from category theory without 
even knowing it -- this is surely the case for many people working 
with Haskell, Purescript and other functional languages. Literate 
programming constitutes a great solution for these developers to learn, by 
providing enough space to introduce categorical concepts 
by example as the code progresses, without forcing 
them -- often used to just skim through READMEs -- to resort 
to external sources to educate themselves.

Note that at the moment not all of our source files are adequately 
presented as {\LaTeX} documents. This is intended, since it is 
desirable to have the source in a stable form before putting massive 
effort towards documenting it.
\section{Some excerpts of the code}
Let us now dive into the code, showing its design principles. We will 
present how we can implement the definition of \emph{category}. We deem this
useful, both to demonstrate code sintax and because such definition 
in Idris may feel very different than the ``traditional'' one, 
at least to someone not used to dependently typed coding. In
\idrisct, we define a category as follows:
\lstset{language=idris, style=numbers}
\begin{lstlisting}
record Category where
  constructor MkCategory
  obj           : Type
  mor           : obj -> obj -> Type
  identity      : (a : obj) -> mor a a
  compose       : (a, b, c : obj)
               -> (f : mor a b)
               -> (g : mor b c)
               -> mor a c
  leftIdentity  : (a, b : obj)
               -> (f : mor a b)
               -> compose a a b (identity a) f = f
  rightIdentity : (a, b : obj)
               -> (f : mor a b)
               -> compose a b b f (identity b) = f
  associativity : (a, b, c, d : obj)
               -> (f : mor a b)
               -> (g : mor b c)
               -> (h : mor c d)
               -> compose a b d f (compose b c d g h)
               	= compose a c d (compose a b c f g) h
\end{lstlisting}
On line $1$, we say that \lstinline|Category| is a \emph{record}, 
that is, a collection of Idris values. To build our record we provide 
a \lstinline|constructor|, which is a function which receives the single
values, as specified in lines $3-21$, and returns the whole data type.
Our record is composed of the fields \lstinline|obj|, 
\lstinline|mor|, \lstinline|identity|, \lstinline|compose|, 
\lstinline|leftIdentity|, \lstinline|rightIdentity| and
\lstinline|associativity|. 
Let us review them in detail.
\begin{itemize}
  \item \lstinline|obj : Type| just says that any type can be taken to be the
  set of objects of our category. The objects of the category will be the 
  terms of that type.
  \item \lstinline|mor : obj -> obj -> Type|, instead, is a
  \emph{function type}. It says that once we specify two objects \lstinline|a|
  and \lstinline|b| of type \lstinline|obj| (standing for the homset source and
  target), then \lstinline|mor a b| is the type of the morphisms going from 
  \lstinline|a| to \lstinline|b|.  This type is not 
  definable in languages that are not dependently typed: The type 
  \lstinline|mor a a| depends on the value \lstinline|a|!

  Note how we opted to define a category by specifying homsets, and 
  not by considering the set of \emph{all} morphsims as a whole and 
  then defining source and target as functions \lstinline|mor -> obj|, as it 
  is done for instance in~\cite{MacLane1978}.
  \item \lstinline|identity| asks for an object \lstinline|a| and produces a 
  term having type \lstinline|mor a a|. This can be thought of as a 
  function that points out which term of type \lstinline|mor a a| has to 
  be considered the identity morphism on \lstinline|a|.
  \item \lstinline|compose| expects objects \lstinline|a, b, c| and 
  two morphsims \lstinline|f: a -> b| and \lstinline|g: b -> c|, and produces 
  a morphism of type \lstinline|mor a c|, so going from \lstinline|a| to 
  \lstinline|c|.

We decided to make types \lstinline|a, b, c| explicit for now because 
it is easier to track the values involved, which is useful for 
experimenting and debugging. These types will be made implicit 
(to reduce bookkeeping) as soon as the code is stable enough.
  \item \lstinline|leftIdentity| is where Idris's dependent type system
  really shines: It expects two objects \lstinline|a| and \lstinline|b| and a morphism 
  \lstinline|f: a -> b|. From this, it produces a term of type 
  \lstinline|compose a a b (identity a) f = f|. A term of this type is a 
  \emph{proof} that the right-hand side and the left-hand sides of that 
  equation are equal, so nothing more than a proof that, for a given 
  morphism \lstinline|f|, composing on the left with the identity 
  morphism of its source amounts to doing nothing.
  Having this kind of terms that represent proofs would be impossible 
  without dependent types.
	\item \lstinline|rightIdentity| is analogous to \lstinline|leftIdentity|.
  \item Finally \lstinline|associativity|, albeit perhaps a bit difficult to 
  parse, implements exactly what its name suggests: It expects three 
  morphisms which can be sequentially composed, and produces a proof 
  that the order of composition does not matter. Hence, a term of type 
  \lstinline|associativity| is just a proof that three fixed morphisms have 
  associative composition.
\end{itemize}
This explanation should help to clarify why we claim that our 
definitions are \emph{verifiably correct}: In \idrisct a term of type 
\lstinline|Category| is not just a collection of objects, identities, 
morphisms and compositions, but it also comes with \emph{proofs} 
that the categorical axioms hold!

To further illustrate this point, \emph{let us prove that Idris types and functions 
form a category in our framework}. This is useful, since it provides a 
bridge between Idris' inner workings and the formal environment we 
are defining. To do this, we will require some helper functions:
\begin{lstlisting}
TypeMorphism : Type -> Type -> Type
TypeMorphism a b = a -> b

identity : (a : Type) -> TypeMorphism a a
identity a = id

compose :
     (a, b, c : Type)
  -> (f : TypeMorphism a b)
  -> (g : TypeMorphism b c)
  -> TypeMorphism a c
compose a b c f g = g . f
\end{lstlisting}
These functions describe the main structure of our category: 
morphisms, identities and compositions. In detail:
\begin{itemize}
  \item \lstinline|TypeMorphism| expects two Idris types 
  \lstinline|a| and \lstinline|b|, and returns the type of all Idris functions 
  from \lstinline|a| to \lstinline|b|.
  \item To define identities, we  just resort to Idris' identity functions, 
  defined for each type.
  \item We proceed by defining morphism composition. Unsurprisingly, 
  we set this to be Idris function composition.
\end{itemize}
It is crucial to note that merely \emph{defining} what the objects, 
morphisms, identities and compositions are is not sufficient in our 
library. We also need to provide \emph{proofs} that the categorical 
laws hold. To accomplish this, we write such proofs as functions:
\begin{lstlisting}[firstnumber=14]
leftIdentity :
     (a, b : Type)
  -> (f : TypeMorphism a b)
  -> f . (identity a) = f
leftIdentity a b f = Refl

rightIdentity :
     (a, b : Type)
  -> (f : TypeMorphism a b)
  -> (identity b) . f = f
rightIdentity a b f = Refl
\end{lstlisting}
As before, \lstinline|discreteLeftIdentity| and 
\lstinline|discreteRightIdentity| are conceptually equal, so we focus 
on the former: We are just using the fact that -- via $\eta$-reduction -- 
Idris considers composing 
a function with an identity equal to the function itself. This is denoted 
in the language using the equality type inhabitant \lstinline|Refl|. We 
implement the associativity law in a very similar way:
\begin{lstlisting}[firstnumber=26]
associativity :
      (a, b, c, d : Type)
  -> (f : TypeMorphism a b)
  -> (g : TypeMorphism b c)
  -> (h : TypeMorphism c d)
  -> (h . g) . f = h . (g . f)
associativity a b c d f g h = Refl
\end{lstlisting}
Here, we rely on the fact that Idris automatically reduces both sides 
to the same normal form, so we can again use \lstinline|Refl|.
Finally, we can put everything together, and obtain:
\begin{lstlisting}[firstnumber=34]
typesAsCategory : Category
typesAsCategory = MkCategory
  Type
  TypeMorphism
  identity
  compose
  leftIdentity
  rightIdentity
  associativity
\end{lstlisting}
Here we are saying that our set of objects is given by the type of all 
types. The morphisms are specified by the type 
\lstinline|TypeMorphism|, while \lstinline|identity| and 
\lstinline|compose| take care of specifying, respectively, what the 
identity morphisms and morphism compositions are. Finally, we plug 
in the laws we implemented above. The category 
\lstinline|typesAsCategory| is important, because it allows us to 
functorially map other categories into actual Idris computations: 
With it, we can formally translate a morphism into a list of functions 
our machine has to compute using functors.
\subsection{What else can you do, and does it scale?}
While the example above may seem a bit limited, \idrisct is far richer 
than this may suggest: We have implemented categories, functors, 
natural transformations, products and coproducts, initial and terminal objects,
monoidal categories, strict monoidal categories and strict symmetric monoidal 
categories. Moreover, there is work in progress, which could be found in the
Github repository, on monads and Kleisli categories, F-algebras and
Eilenberg-Moore categories, wiring diagrams of discrete dynamical
systems. Curiously, in implementing non-strict, symmetric 
monoidal categories we seem to have pushed the Idris compiler to its 
limits~\cite{Perone2019}, and we still need to figure out a way to 
help it recognize that our code should indeed typecheck.

We also have provided some useful instances of categories, e.g. we 
proved that categories and functors themselves -- that is, terms of 
type \lstinline|Category| and \lstinline|Functor| in our 
implementation -- form a category. This means that we have 
implemented $\textbf{Cat}$ in \idrisct. We also proved that monoids 
are categories, and that they can be used to \emph{freely generate} the 
objects of a monoidal category. Having done so, we are now focused 
on implementing free categories, such as the free strict symmetric 
monoidal category generated by a set of objects and a set of morphisms.

As one can see, at the moment we are focusing on a narrow subfield 
of category theory: Basic definitions, monoidal categories and limits. 
This is dictated by internal business reasons, but we hope that the 
library will be expanded in different directions with the help of the 
community (in fact, we already received external pull requests defining 
opcategories, cocones, and dualizing our definitions of product and terminal object).

As for scalability, up to now we mainly found two types of blockers: 
In defining free structures, we often had to rely on quotients, 
which are not easily implemented in Idris. This is a well known 
problem which depends on how dependently typed languages deal 
with equality, which could be circumvented as explained in~\cite{Licata2011}.
In some circumstances -- namely when a quotient-free 
way of defining the structure we want to implement exists -- this issue 
can be circumvented by just implementing the quotient-free construction 
as it is. This is the case, for instance, of \emph{free categories}, which 
can be implemented without relying on quotiens by using the notion of 
\emph{path category}.~\cite{nCatLab}

Another source of problems arises when trying to tie our categorical 
definitions with Idris' internal structure. For example, we want to 
prove that Idris types and functions do not just form a category, 
but a monoidal one where the usual product defines the tensor. Since 
our definition of monoidal category is quite verbose, doing this 
directly is difficult. Instead, we opted for proving that products and 
terminal objects define a monoidal structure in the general case, and 
then use this proof to implement our claim. In practice, this required 
developing tools to comfortably deal with commutative diagrams 
and diagram chasing, since dealing with commutative diagrams 
na\"ively makes everything untractable fairly quickly.

Moreover, when dealing directly with Idris functions, 
sometimes we were forced to assume function extensionality.
Natural transformations, for example, are considered equal if 
their families of morphisms are the same. These families of 
morphisms are modelled as functions, so extensionality is 
necessary to arrive at the correct notion of equality.
This is not a problem per sé since Idris cannot internally distinguish between 
intensional and extensional functions, but we deemed it worth noticing.
\section{Open problems and future work}
Future work includes extending \idrisct via external contributions. 
Indeed, raising awareness about this project within the research 
community is the main reason for this submission. Figuring 
out which companies focused on using applied category theory in 
developing products, and thinking about ways that could speed 
adoption of category theory into industry were a big topic of discussion 
at ACT 2018.  We decided to make the \idrisct library open source to 
help this process, and also as an act of gratitude towards the whole 
ACT community, which helped us figuring out things many times. 
We would be very happy, then, if other researchers from the ACT 
community would converge on \idrisct, by opening pull requests 
or by forking the repository, so that the library could be extended beyond 
the use that Statebox wants to make of it.

On a different note, many type definitions contain arguments that 
will have to be made implicit. This will require code refactoring, 
which will surely happen in the future. At the moment we are 
postponing this task since the conversion of arguments to implicits 
will result in less bookkeeping (which is good), but it will also make 
compiler debugging trickier and compilation longer (which is bad). 
Hence, the team is waiting for the code to become stable and 
crystallized enough before undertaking such task.

Another direction of future work includes interfacing the \idrisct 
library with \texttt{typedefs}~\cite{StateboxTeam2018a}, the other big 
Idris project the team is pursuing at the moment. 
\emph{\texttt{typedefs} is a language-agnostic type construction 
language based on polynomials}, that makes it possible to pass types 
and terms around between different programming languages with ease. 
Incredibly, as of now (April 2019), \texttt{typedefs} is the most 
trending Idris repository on Github, a clear signal that the functional 
programming community considers it to be a useful project. In our 
long-term vision, developers will be able to use \texttt{typedefs} to 
interface the non-core parts of their projects, such as the user 
interface, with the core parts -- written in Idris using \idrisct -- where 
the verified implementation of some categorical gadget will reside. 
This will allow for an easy passing of data between formally consistent 
environments, where core evaluations are carried out, and more 
flexible environments, where the frontend material is provided. In 
practice, this requires to prove that \texttt{typedefs} is a monoidal 
category, in a similar fashion to what we did with Idris types and 
functions.

Finally, it has to be considered that the library is still a prototype: It may 
require tweaking and optimization in order to become fully 
performant. However, this will also have to wait until the codebase 
is in a more settled state. Another important consideration in this 
regard is the status of the successor to Idris, which has been tentatively 
named Blodwen. It may make sense to postpone optimization efforts 
only after it has been released and become viable.

\section*{Acknowledgements}
We want to thank: Our fellow team members at Statebox and the friendly, 
stimulating environment that Statebox itself provides, which made this contribution
possible; our reviewers, which provided valuable advice and observations which
we did our best to incorporate in the paper; the whole ACT community, which is always welcoming and supporting.

Finally we want to thank the unsung hero of our library, Github user 
\texttt{mstn} (unfortunately we do not know his/her real identity), 
that keeps enriching \idrisct day after day with valuable commits 
and pull requests.
\printbibliography
\end{document}